\documentclass[10pt]{forma}
\usepackage{graphicx}
\def\be{\begin{equation}}
\def\ee{\end{equation}}

\def\bea{\begin{eqnarray}}
\def\eea{\end{eqnarray}}

\begin{document}
\hspace{1.0cm} \parbox{15.0cm}{

\baselineskip = 15pt

\noindent {\bf DARK ENERGY AND LARGE-SCALE STRUCTURE OF THE UNIVERSE}
\bigskip
\bigskip

\noindent {\bf Yu. Kulinich, B. Novosyadlyj}
\bigskip
\bigskip

\baselineskip = 9.5pt

\noindent {\small {\it Astronomical Observatory, Ivan Franko National University of Lviv\\
8 Kyryla \& Methodia Str., 79005 Lviv, Ukraine}} \\
\noindent {\small {\it e-mail:}} {\tt kul@astro.franko.lviv.ua}

\smallskip
\baselineskip = 9.5pt \medskip

\medskip \hrule \medskip

\noindent 

The evolution of matter density perturbations  in two-component model of the Universe consisting of dark energy (DE) and dust-like matter (M) is considered. We have analyzed it for two kinds of DE with $\omega\ne -1$: a) unperturbed energy density and b) perturbed one (uncoupled with matter). For these cases the linear equations for evolution of the gauge-invariant amplitudes of matter density perturbations are presented. It is shown that in the case of unperturbed energy density of DE the amplitude of matter density perturbations grow slightly faster than in the second case.

\medskip \hrule \medskip

}
\vspace{1.0cm}

\renewcommand{\thefootnote}{\ }

\footnotetext{\copyright~Yu. Kulinich, B. Novosyadlyj, 2004}

\renewcommand{\thefootnote}{\arabic{footnote}}

\baselineskip = 11.2pt
\noindent {\small {\bf INTRODUCTION}}
\medskip

 The measurements of luminosity distances $d_L$ to the SN Ia stars as function of redshift have revealed the accelerated expansion of the Universe \cite{perlmutter99,riess98}. Until recently for the interpretation of collected data on large scale structure of the Universe models with positive cosmological constant $\Lambda$ were preferred by cosmologists. For such models one can calculate the evolution of matter density perturbations up to formation of gravitationally bound systems of galaxies and clusters of galaxies (see \cite{knov} and references therein). Searching of plausible physical interpretation of $\Lambda$-constant has introduced into the astrophysics new terms: dark energy (DE) and quintessence for the notation of energy of unknown nature that repulses and involves the self-attracting matter into accelerated expansion. Classical $\Lambda$-constant is the simplest kind of such energy. Now the more general models of this component are under considerations (see for review \cite{peebls} and references therein). In some papers the assumption of absence of coupling between DE and matter is used. But in this case the matter density perturbations lead to perturbations of dark energy density \cite{ma,dave,doran}. Other kind of DE is based on the assumption of homogeneous and isotropic distribution of this component. Such models predict energy flow from one component to another or in other words DE and matter are coupled in perturbed regions. In this paper we will analyze the evolution of matter density perturbations for both kinds of DE with constant equation of state and $\omega^{(DE)}=P^{(DE)}/\varepsilon^{(DE)}\ne -1$, where $P^{(DE)}$ and $\varepsilon^{(DE)}$ are pressure and energy density of DE respectively.

\medskip
\noindent {\small {\bf DARK ENERGY}}
\medskip

The influence of DE upon dynamics of the Universe and evolution of matter density perturbations can be studied by analysis of Einstein equations and its presentation as ideal fluid with equation of state $P=\omega \varepsilon$, where $\omega$ is negative. In the case of  $\omega=-1$ we have $\Lambda$-constant or Lorentz-invariant dark energy which  can be  presented also by density of Lagrangian function
$\mathcal{L} \equiv \mathcal{L}(\{g_{ik}\},\{\frac{\partial g_{ik}}{\partial x^l}\})$
that satisfies the equation
$$
\frac12\sqrt{-g}\Lambda g_{ik} = \frac{\partial (\sqrt{-g}\mathcal{L})}{\partial g^{ik}} -
\frac{\partial}{\partial x^l}\frac{\partial (\sqrt{-g}\mathcal{L})}{\partial \frac{\partial g^{ik}}{\partial x^l}},
$$
where $\Lambda \equiv \Lambda(\{g_{ik}\},\{\frac{\partial g_{ik}}{\partial x^l}\})$ is some arbitrary function. The Einstein $\Lambda$-constant has the physical interpretations of 
zero-point vacuum fluctuations, vacuum polarization or follows from some versions of supersymmetry theories.
But all these interpretations converge to the fine tuning problem: at Planckian epoch the energy 
density of matter was by $\sim 120$ orders larger than dark energy one.
This issue can be essentially relaxed when $\omega\ne -1$ and depends on time, this called dark energy of tracker field kind. For the such type of DE besides of the state equation and coupling of DE with matter we should define the vector of 4-velocity ($\vec u$, $(\vec u)^2=-1$) which indicate the direction of energy flow. Using this vector we can define the 3d metrics tensor  $h_{ik} = u_iu_k -g_{ik}$ in 4d space-time and thermodynamic parameters such as the {\it energy density} $\varepsilon$, {\it pressure density} $P$ and {\it anisotropic stress-tensor} $\Pi_{ik}$:
\begin{eqnarray}
\varepsilon = T_{ik}u^iu^k,\quad
P = \frac13 T_{ik}h^{ik},\quad
\Pi_{ik} = T_{jl}\left(h^j_i h^l_k -\frac13h^{jl} h_{ik}\right).\nonumber
\end{eqnarray}
In cosmological applications the constant equation of state $P=\omega\varepsilon$ with $\omega=0$ for the case of dust-like pressureless matter, $\omega=1/3$ for electromagnetic field and $\omega=-1$ for the case of Einstein's $\Lambda$-constant are used.
For the case of scalar fields the general form of $\omega=\omega(\tau)$ may exist.
The scalar field has a density of Lagrangian:
$\mathcal{L} = \frac12 g^{ij}\varphi_{,i}\varphi_{,j} - V(\varphi),$ leading to 
$\varepsilon^{(\varphi)} = \frac12 {\dot \varphi }^2 + V(\varphi)$ and 
$P^{(\varphi)} = \frac12 {\dot \varphi }^2 - V(\varphi)$
for $\varphi$ homogeneously and isotropically distributed on 3d hypersurface.
The DE of such kind finds its interpretation in the framework of generalizations of gravitation theory, e.g. branes theory and theory of gravitation with more general geometry than Riemann's one or some unified theories of fundamental physical interactions.

\medskip  \medskip
\noindent {\small {\bf MATTER AND DARK ENERGY}}
\medskip

For two-component model of Universe consisting of matter and DE the energy-stress tensor is represented by
$T^i_k = T^{(M)}{}^i_k + T^{(DE)}{}^i_k.$
Conservation equations are written as
$\nabla_i \left( T^{(M)}{}^i_k + T^{(DE)}{}^i_k \right) = 0,$
or in other form $\nabla_i T^{(M)}{}^i_k = Q_k$ and $\nabla_i T^{(DE)}{}^i_k = - Q_k$,
where $\vec{Q}\equiv\{Q_k\}$ is vector of energy flow \cite{kodama}. For the general case one should define the vector of energy flow between two components.

We assume unperturbed DE and comoving matter and DE on the homogeneous and isotropic background. This simplification leads to $\overline{\nabla}_i T^{(DE)}{}^i_{k} = 0$,
where $\overline{\nabla}_i$ is covariant derivative in isotropic and homogeneous
space with metrics tensor $\overline{g}_{ik}$. The real perturbed space-time presented by metrics $g_{ik}$. The motion equations for particles comoving to unperturbed
background and DE (in the case of $\overline{\Pi}_{ik}^{(DE)}=0$) are following
\be
\frac{d\overline{u}^i}{ds} = - \overline{\Gamma}^i_{jk}\overline{u}^j\overline{u}^k
= - \Gamma^i_{jk}\overline{u}^j\overline{u}^k + f^i,\nonumber
\ee
where $ \overline{\Gamma}^i_{jk}$ and  $\Gamma^i_{jk}$ are Christoffel symbols defined for the metrics $\overline{g}_{ik}$ and $g_{ik}$ respectively, $f^i = (\Gamma^i_{jk}-\overline{\Gamma}^i_{jl})\overline{u}^j\overline{u}^l$ is vector of additional force needed for the homogeneous distribution of the DE in the regions of perturbations. The vector of energy flow is $Q_k = \overline{\nabla}_i T^{(DE)}{}^i_{k} - \nabla_i T^{(DE)}{}^i_{k}$, that gives
$$
Q_k = W^j_{kl} T^{(DE)}{}^l_j-W^l_{jl} T^{(DE)}{}^j_k,
$$
where tensor $W^i_{jk} \equiv \Gamma^i_{jk}-\overline{\Gamma}^i_{jk}$. Definition of perturbations is ambiguous and depends on a choice of gauge.

The metrics in the longitudinal gauge has a form
$ds^2 = a(\eta)^2[-(1+2\Psi (\eta) Y(x^{\alpha}))d\eta^2 + (1+2\Phi (\eta) Y(x^{\alpha}))\delta_{\beta\gamma}dx^{\beta}dx^{\gamma}]$ where $\Phi(\eta)$ and $\Psi(\eta)$ are Bardeen's potentials \cite{bardeen}. Non-zero components of tensor $W^i_{jk}$ according to this metrics are ($\alpha,\beta=1,2,3$)
\bea
W^{0}_{00} = \dot \Psi Y,\qquad
W^{0}_{0 \alpha} = W^{0}_{\alpha 0} = -k\Psi Y_{\alpha},\qquad
W^{\alpha}_{00} = -k\Psi Y^{\alpha},\nonumber\\
W^{\beta}_{0 \beta} = W^{\beta}_{\beta 0} = \dot \Phi Y,\qquad
W^{0}_{\beta\beta} = \left(2\frac{\dot a}{a}(\Phi-\Psi)+\dot \Phi \right)Y,\nonumber\\
W^{\beta}_{\beta \alpha} = W^{\beta}_{\alpha\beta} = -k\Phi Y_{\alpha}\quad \mbox{and} \quad
W^{\alpha}_{\beta \beta} = k \Phi Y^{\alpha} \quad \mbox{for } \alpha\ne \beta\; ({\rm no\; sum\; on}\; \beta).\nonumber
\eea
Thus the components of vector of energy flow and additional force are
\bea{}
Q_0 &=& 3\dot\Phi (\varepsilon^{(DE)}+P^{(DE)})Y, \quad Q_{\alpha} = k\Psi (\varepsilon^{(DE)}+P^{(DE)})Y_{\alpha},\label{Q} \\
{\rm and} &&\nonumber \\
f^0 &=& \dot\Psi Y, \quad  f^{\alpha} = -k\Psi Y^{\alpha}\label{f}
\eea
respectively.

\medskip  \medskip
\noindent {\small {\bf EVOLUTION OF MATTER DENSITY PERTURBATIONS}}
\medskip

Formation of the large scale structure of the Universe is described by linear theory of scalar perturbations. We use here the gauge-invariant approach presented in \cite{bardeen, root, kodama, mukhanov}. We have considered a case of two-components Universe with small perturbations in a component of the dust-like matter and no perturbations in a component of the DE. The non-zero components of energy-stress tensors for every component are
\bea{}
T^{(M)}{}^0_0 = -\overline{\varepsilon}^{(M)}(1+\delta^{(M)}Y),\quad
T^{(M)}{}^0_{\alpha} = (\overline{\varepsilon}^{(M)}+ \overline{P}^{(M)})vY_{\alpha},\quad
T^{(M)}{}^{\alpha}_0 = (\overline{\varepsilon}^{(M)}+ \overline{P}^{(M)})vY^{\alpha}, \nonumber \\
T^{(M)}{}^{\alpha}_{\beta} = \overline{P}^{(M)}
[(1+\pi^{(M)}Y)\delta^{\alpha}_{\beta}+\Pi^{(M)} Y^{\alpha}_{\beta}],\quad
T^{(DE)}{}^0_0 = -\overline{\varepsilon}^{(DE)},\qquad
T^{(DE)}{}^{\alpha}_{\beta} = \overline{P}^{(DE)}\delta^{\alpha}_{\beta}, \nonumber
\eea{}
where $\delta$ and $v$ are perturbations of energy density and velocity respectively,
$\pi^{(M)}$ and $\Pi^{(M)}$ isotropic and anisotropic components of pressure perturbations
(over-lines denote the background magnitudes). From Einstein's equations $\delta G^0_0 = 4\pi G \delta T^0_0$, $\delta G^0_{\alpha} = 4\pi G \delta T^0_{\alpha}$ and $\delta G^{\beta}_{\alpha} = 4\pi G \delta T^{\beta}_{\alpha}$ we obtain the following connection between perturbations of metrics and perturbations of matter density and velocity:
\bea
4\pi G a^2 \overline{\varepsilon}^{(M)} D^{(M)} &=& (k^2 - 3K)\Phi, \label{aeq1}\\
4\pi G a^2 (\overline{\varepsilon}^{(M)} + \overline{P}^{(M)})V^{(M)} &=& k\left(\left(\frac{\dot a}{a}\right)\Psi - \dot \Phi\right),\label{aeq2}\\
4\pi G a^2 \overline{P}^{(M)} \Pi^{(M)} &=& -k^2(\Phi+\Psi).\label{aeq3}
\eea
The conservation equations $\delta T^{(M)}{}_0^i{}_{|i}=Q_0$ and $\delta T^{(M)}{}_{\alpha}^i{}_{|i}=Q_{\alpha}$ lead to the following  equations for matter density and velocity perturbations:
\bea
\dot D_g^{(M)} + 3({c^{(M)}_s}^2-\omega^{(M)})\frac{\dot a}{a}D_g^{(M)} + (1+\omega^{(M)})kV^{(M)} + 3\omega^{(M)}\frac{\dot a}{a}\Gamma^{(M)} =
-3\dot \Phi \frac{\varepsilon^{(DE)}}{\varepsilon^{(M)}}(1+\omega^{(DE)})\nonumber\\
\dot V^{(M)} + \frac{\dot a}{a}(1-3{c^{(M)}_s}^2)V^{(M)} - k(\Psi - 3{c^{(M)}_s}^2\Phi) - \frac{{c^{(M)}_s}^2k}{1+\omega^{(M)}}D_g^{(M)} - \frac{\omega^{(M)} k}{1+\omega^{(M)}}\left[\Gamma^{(M)} -\frac32\left(1-\frac{3K}{k^2}\right)\Pi^{(M)}\right] = \nonumber\\
k\Psi\frac{\varepsilon^{(DE)}}{\varepsilon^{(M)}} \frac{1+\omega^{(DE)}}{1+\omega^{(M)}},
\nonumber
\eea
where $V=v$, $D_g=\delta+3(1+\omega)\Phi$, $D=\delta+3(1+\omega)\frac{\dot a}{a}\frac{V}{k}$, $\Gamma=\pi-\frac{c_s^2}{\omega}\delta$ are gauge-invariant amplitudes \cite{bardeen}-\cite{mukhanov} ($c_s^2 = \dot P/\dot \varepsilon$
is square of sound speed).

For DE component we have $\delta T^{(DE)}{}_0^i{}_{|i}=-Q_0$ and
$\delta T^{(DE)}{}_{\alpha}^i{}_{|i}=-Q_{\alpha}$ that gives the equations $\dot D^{(DE)} = 0$ and $\dot V^{(DE)} = 0$. If we suppose
initial zero perturbations of dark energy ($D_{in}^{(DE)}=0$, $\Gamma_{in}^{(DE)}=0$ and $V_{in}^{(DE)}=0$) then $D^{(DE)}=V^{(DE)}=0$.

For dust-like matter $\Pi^{(M)}=c_s^{(M)}=\omega^{(M)}=\Gamma^{(M)}=0$ the conservation equations are simplified:
\bea
\dot D_g^{(M)} + kV^{(M)}  = -3\dot \Phi\frac{\varepsilon^{(DE)}}{\varepsilon^{(M)}}(1+\omega^{(DE)}),\quad
\dot V^{(M)} + \frac{\dot a}{a}V^{(M)} - k\Psi = k\Psi\frac{\varepsilon^{(DE)}}{\varepsilon^{(M)}} (1+\omega^{(DE)}).
\label{cons_dm}
\eea
The set of equations (\ref{aeq1})-(\ref{aeq3}) and (\ref{cons_dm}) describes the evolution of scalar perturbations of matter in the Universe with DE which is homogeneously distributed over whole space.
For the case of DE uncoupled with dust-like matter the right-hand sides of the equations (\ref{cons_dm}) will be equal zero. In this case the energy density of DE will trace matter density perturbations by means of metrics perturbations, so, it will be perturbed \cite{ma,dave,doran}. The evolution of perturbation amplitudes of matter density $D_g^{(M)}$  for the two cases of DE with unperturbed energy density and perturbed one are shown in Fig.1. For calculation the following values of parameters have been used: the constant state equation parameter of DE is $\omega^{(DE)}=-0.8$, the current contents of DE and matter are respectively $\Omega^{(DE)}=0.7$ and $\Omega^{(M)}=0.3$, and dimensionless Hubble constant is $h=0.65$. The amplitude $D_g^{(M)}$ is larger for case of unperturbed DE that is stipulated by flow of energy from DE component to matter one in perturbed region.

\begin{figure}[htbp]
\centerline{\includegraphics[width=9.0cm]{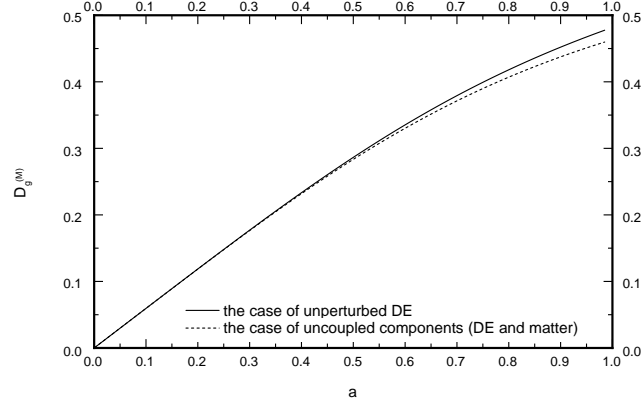}}
\caption{The dependence of $D_g^{(M)}$ on scale factor $a$ for two kinds of DE:
uncoupled with dust-like matter (dashed line) and unperturbed DE (solid line).
The state of DE is $\omega^{(DE)}=-0.8$ and other cosmological parameters are
$\Omega^{(DE)}=0.7$, $\Omega^{(M)}=0.3$, $k=10^{-1}$ Mpc${}^{-1}$, $h=0.65$.}
\label{grape}
\end{figure}

\medskip  \medskip
\noindent {\small {\bf CONCLUSIONS}}
\medskip

We have analyzed the evolution of matter density perturbations for two kinds of dark energy:
unperturbed homogeneously distributed  and uncoupled with dust-like matter. The expressions for energy flow between components (\ref{Q}) and additional force which keeps homogeneous distribution of DE (\ref{f})
as well as the equations for evolution of matter density perturbations (\ref{aeq1})-(\ref{aeq3}) and (\ref{cons_dm}) are obtained. Their numerical solutions show that gauge-invariant amplitude of matter density perturbations grow slightly faster in the case of homogeneously distributed DE than in the case of dark energy uncoupled with matter. This difference can be explained by existence of flow of energy from DE to dust-like matter and additional force smoothing the DE.



\end{document}